\documentclass[12pt]{article}
\usepackage{geometry} % see geometry.pdf on how to lay out the page. There's lots.
\usepackage[english]{babel}
\usepackage[english]{layout}

\usepackage{mathpazo} % math & rm
\usepackage[scaled]{helvet} % ss
\usepackage{courier} % tt
\normalfont
\usepackage[T1]{fontenc}

\usepackage{array} % for better arrays (eg matrices) in maths
\usepackage{paralist} % very flexible & customisable lists (eg. enumerate/itemize, etc.)
\usepackage{verbatim} % adds environment for commenting out blocks of text & for better verbatim
\usepackage[utf8]{inputenc} % Any characters can be typed directly from the keyboard, eg éçñ
\usepackage{textcomp} % provide lots of new symbols
\usepackage{graphicx}  % Add graphics capabilities
\usepackage{epstopdf} % to include .eps graphics files with pdfLaTeX
\usepackage{flafter}  % Don't place floats before their definition

\usepackage{amsfonts,latexsym,amsthm,amssymb}
\usepackage{amsmath,amssymb,bbm,amsfonts}  % Better maths support & more symbols
\usepackage{bm}  % Define \bm{} to use bold math fonts

\usepackage{memhfixc}  % remove conflict between the memoir class & hyperref
\usepackage{fancyhdr} % This should be set AFTER setting up the page geometry
\usepackage{sectsty}
\allsectionsfont{\sffamily\mdseries\upshape} % (See the fntguide.pdf for font help)
\numberwithin{equation}{section}

\def\Ref{\ref}
\def\Ref#1{(\ref{#1})}

\def\eb{\begin{eqnarray*}}
\def\ee{\end{eqnarray*}}
\newcommand{\bc}{\begin{center}}
\newcommand{\ec}{\end{center}}

\title{Jordanian deformation of the open XXX-spin chain}
\author{ \textsf{P. ~P.~Kulish,}
\thanks{E-mail address: kulish@euclid.pdmi.ras.ru}
\textsf{ ~~N. ~Manojlovi\'c}
\thanks{E-mail address: nmanoj@ualg.pt}
\textsf{ and Z. ~Nagy}
\thanks{E-mail address: 
zoltan.nagy@m4x.org} \\
\\
\textit{$^{\ast}$St. Petersburg Department of Steklov Mathematical Institute} \\
\textit{Fontanka 27, 191023, St. Petersburg, Russia} \\
\\
\textit{$^{\ast \dag\ddag}$
Grupo de F\'{\i}sica Matem\'atica da Universidade de Lisboa} \\
\textit{Av. Prof. Gama Pinto 2, PT-1649-003 Lisboa, Portugal} \\
\\
\textit{$^{\dag}$Departamento de Matem\'atica, F. C. T.,
Universidade do Algarve}\\ 
\textit{Campus de Gambelas, PT-8005-139 Faro, Portugal}}
\date{}

\begin{document}

\maketitle
\thispagestyle{empty}
\abstract{The general solution to the reflection equation associated with the jordanian deformation of the $SL(2)$ invariant Yang R-matrix is found. The same K-matrix is obtained by the special scaling limit of the XXZ-model with general boundary conditions. The Hamiltonian with the boundary terms is explicitly derived according to the Sklyanin formalism. We discuss the structure of the spectrum of the deformed XXX-model and its dependence on the boundary conditions.

}

\clearpage 
\newpage

%
%%%%%%%%%%% Introduction %%%%%%%%%%%
%

\section{Introduction}
%The theory of quantum groups (a type of Hopf algebra) arose from the quantum inverse scattering method approach to the resolution and construction of quantum integrable systems \cite{Faddeev}. 

The quantum inverse scattering method (QISM) \cite{TakhFadI, SklyTakhFad, Faddeev, KulSkl} as an approach to construct and solve quantum integrable systems has lead to the foundations of the theory of quantum groups \cite{Drinfeld,Jimbo}. A particularly interesting feature of quantum groups is that by a transformation called \emph{twist} \cite{DrinfeldT}, one can create new quantum groups starting from known ones. Although the twist transformations generate equivalence relation among quantum groups, they produce different R-matrices. These new  R-matrices, in turn, can lead to new integrable systems \cite{Kul1}.

Twist of a quantum group, or more general Hopf algebra $\mathcal{A}$, is a similarity transformation of the coproduct $\Delta: \mathcal{A}\rightarrow \mathcal{A} \otimes \mathcal{A}$ by an invertible twist element $\mathcal{F} = \sum _{j} f_j^{(1)}\otimes f_j^{(2)} \in \mathcal{A}\otimes \mathcal{A}$,
\begin{equation}
\label{twDelta}
\Delta (a) \to \Delta_{t} (a) = \mathcal{F} \Delta (a) \mathcal{F}^{-1}, \quad a \in \mathcal{A}.
\end{equation}
In order to guarantee the coassociativity property of the coproduct, the element $\mathcal{F}$ has to satisfy certain compatibility condition, the so-called twist equation \cite{DrinfeldT}
\begin{equation}
\label{TwEq}
\mathcal{F}_{12} \left( \Delta \otimes \mathrm{id} \right) \mathcal{F} = \mathcal{F}_{23}\left(\mathrm{id} \otimes  \Delta\right) \mathcal{F} ,
\end{equation}
where $\left( \Delta \otimes \mathrm{id} \right) \sum _{j} f_j^{(1)} \otimes f_j^{(2)}= \sum _{j} \Delta\left(f_j^{(1)}\right) \otimes f_j^{(2)} \in \mathcal{A}\otimes \mathcal{A}\otimes \mathcal{A}$. 
Moreover, the transformation law of the coproduct also determines how the corresponding universal $R$-matrix is changed
\begin{equation}
\label{twtR}
\mathcal{R} \to \mathcal{R} ^{(t)} = \mathcal{F}_{21} \mathcal{R} \mathcal{F}^{-1}, \quad \mathcal{F}_{21} = \sum _{j} f_j^{(2)} \otimes  f_j^{(1)}.
\end{equation}
This new $R$-matrix allows us to build and study new integrable models \cite{KulSto1}.

A particular solution of the twist equation is provided by the jordanian twist element for the enveloping algebra of the $sl(2)$ Lie algebra which appeared in \cite{Ge,Og} and it was extended to the $sl(N)$ case in \cite{Kul1,KuLyMu}. In this paper we consider the example of the twisted algebra $sl_\theta(2)$ and the corresponding deformation of the Yangian $\mathcal{Y}(sl(2))$ \cite{KulSto1,KhorStolTol}. Since the twist preserves the regularity property of the $R$-matrix (i.e. $R(0)=\eta \mathcal{P}$ where $\mathcal{P}$ is the permutation map of the neighbouring spaces), one can write the deformed version of the integrable Heisenberg XXX spin chain with periodic boundary conditions \cite{KulSto1}
\begin{equation}
%\label{ }
H=\sum_{j=1}^{N} \left(\frac{1}{2}\left(\sigma_j^x\sigma_{j+1}^x+\sigma_j^y \sigma_{j+1}^y +\sigma_j^z \sigma_{j+1}^z\right)+\theta \left(\sigma_j^+-\sigma_{j+1}^+\right)+\theta^2 \sigma_j^+\sigma_{j+1}^+ \right). \notag
\end{equation}

Note that this operator is non-hermitian which gives rise to additional difficulties in the application of the algebraic Bethe ansatz to this model. Although it can be seen that the extra terms added to the $XXX$ Hamiltonian do not change the spectrum of the model \cite{KulSto1,He}, the explicit form of the Bethe states is less straightforward.

In this paper we study the deformation by the jordanian twist of the $XXX$-model with non-periodic boundary conditions. 
The latter ones are described by the reflection matrices $K^{\pm}(\lambda)$ \cite{Sk}. Our result is the classification of reflection matrices compatible with the twisted jordanian $R$-matrix. The general solution of the reflection equation is obtained by a direct calculation and it is also confirmed by the singular scaling limit from the know reflection matrix of the anisotropic XXZ-model. Using the general solution for $K(\lambda)$ and following Sklyanin approach \cite{Sk}, we construct the Hamiltonian with the general non-periodic boundary conditions. We conclude by some remarks on the influence of the boundary conditions on the spectrum of the system.

%\clearpage 
%\newpage

%
%%%%%%%%%%% Solutions of the reflection equation %%%%%%%%%%%
%

\section{Solutions of the reflection equation}

The main relation of the quantum inverse scattering method \cite{TakhFadI, Faddeev}
\begin{equation}
\label{RTT}
R_{12}(\lambda-\mu) T_1(\lambda)T_2(\mu) = T_2(\mu) T_1(\lambda)R_{12}(\lambda-\mu)
\end{equation}
can be used to define a special infinite dimensional quantum algebra - the Yangian. The Yang R-matrix 
\begin{equation}
\label{YangR}
R_{12}(\lambda) = \lambda \mathrm{I} + \eta \mathcal{P} \in \mathrm{End} \left( \mathbb{C}^n \otimes \mathbb{C}^n \right)
\end{equation}
gives rise to the Yangian $\mathcal{Y}(sl(n))$ with the entries of the $n\times n$ matrix $T(\lambda)$ as generating functions of the $\mathcal{Y}(sl(n))$ generators (cf. \cite{CharyP}). We use $\mathrm{I}$ and $\mathcal{P}$ for the identity operator and the permutation in $\mathbb{C}^n \otimes \mathbb{C}^n$: $\mathcal{P} (v\otimes w) = w\otimes v$; $v,w \in \mathbb{C}^n$ and the standard notation of the QISM: $T_1=T\otimes \mathrm{I}, \ T_2= \mathrm{I} \otimes T$. 
The Heisenberg XXX-spin chain is related to the $\mathcal{Y}(sl(2))$, and the universal enveloping algebra of $sl(2)$ is a Hopf subalgebra of the Yangian: $U(sl(2)) \subset \mathcal{Y}(sl(2))$. The two generators $h$ and $X^+$ of $sl(2)$
\begin{equation}
\label{sl2}
[h , X^{\pm}] = \pm 2 X^{\pm} , \qquad [X^+,X^-] = h ,
\end{equation}
give rise to a jordanian twist element
\begin{equation}
\label{jF}
\mathcal{F} = \exp (h\otimes \ln (1+\theta X^+)) \in U(sl(2)) \otimes U(sl(2))
\end{equation} 
which satisfies the Drinfeld twist equation \Ref{TwEq}. The matrix form of $\mathcal{F}$ in the spin {\textonehalf}  representation $\rho$ is $F_{12}\in \mathrm{End} \left(\mathbb{C}^2 \otimes \mathbb{C}^2 \right)$
\begin{align}
   \label{jFonehalf}
F_{12} &= \left( \rho \otimes \rho \right) \mathcal{F} = \exp \left( \sigma ^z \otimes \theta \sigma ^+ \right) = \mathbbm{1} + \theta \sigma ^z \otimes \sigma ^+ \notag  \\
    & = \left(\begin{array}{rrrr}1 & \theta & 0 & 0 \\0 & 1 & 0 & 0 \\0 & 0 & 1 & -\theta \\0 & 0 & 0 & 1\end{array}\right) ,
\end{align}
where $\sigma ^z,  \sigma ^{\pm} = (\sigma ^x \pm \sigma ^y)/2$ are Pauli sigma matrices.

Hence, the $R$-matrix of the twisted Yangian $\mathcal{Y}_\theta(sl(2))$ has the following form \cite{KulSto1}
\begin{equation}
\label{Rmat}
R^{(j)}(\lambda)=F_{21}R_{12}(\lambda) F_{12}^{-1}= \lambda R^{(j)}+ \eta \mathcal{P}  =
\left(
\begin{array}{cccc}
  \lambda+\eta & -\lambda \theta & \lambda \theta & \lambda \theta^2 \\
  0 & \lambda & \eta & - \lambda \theta \\
  0 & \eta & \lambda & \lambda \theta \\
  0 & 0 & 0 & \lambda +\eta
\end{array}
\right) ,
\end{equation}
where $F_{21} = \mathcal{P} F_{12}\mathcal{P}$. This R-matrix is also a solution of the Yang-Baxter equation
\begin{equation}
\label{YBE}
R_{12}(\lambda -\mu) R_{13}(\lambda) R_{23}(\mu)=R_{23}(\mu) R_{13}(\lambda) R_{12}(\lambda-\mu).
\end{equation}

The unitarity property of the $R$-matrix is unaffected by the twist 
\begin{equation}
R_{12}(\lambda)R_{21}(-\lambda)= g(\lambda) ,
\end{equation}
with $g(\lambda)=(-\lambda ^2+\eta^2)$. But its PT symmetry is broken
\begin{equation}
R_{21}(\lambda)\neq R_{12}(\lambda)^{t_1t_2},
\end{equation}
here $R_{21}(\lambda) = \mathcal{P} R_{12}(\lambda) \mathcal{P}$, and the indices $t_1 , t_2$ denote the transpositions in the first and in the second space of the tensor product $\mathbb{C}^2 \otimes \mathbb{C}^2$, correspondingly.
The $R$-matrix does not have the crossing symmetry either, but it satisfies a weaker condition
\begin{equation}
\label{weakcross}
\{\{\{R_{12}(\lambda)^{t_2}\}^{-1}\}^{t_2}\}^{-1}=\frac{g(\lambda+\eta)}{g(\lambda+2\eta)}M_2 R_{12}(\lambda+2\eta)M_2^{-1} ,
\end{equation}
with a matrix 
\begin{equation}
\label{Mmat}
M=\left(\begin{array}{cc} 1 & -2\theta\\ 0 & 1 \end{array} \right) .
\end{equation}
One can point out that a more general matrix 
\begin{equation}
\label{Mgen}
\widetilde{M}=\left(\begin{array}{cc} 1 & \alpha \\ 0 & 1 \end{array} \right) 
\end{equation}
commutes with the $R$-matrix
\begin{equation}
\label{Mcom}
\left[ \widetilde{M} \otimes \widetilde{M} , R(\lambda) \right]=0 .
\end{equation}
 
As a remark on general level we can say that for our purposes it is enough that the matrix $\{\{\{R_{12}(\lambda)^{t_2}\}^{-1}\}^{t_2}\}^{-1}$ \emph{exists}. The fact that this matrix can be put in the form \eqref{weakcross} only implies that we can establish a bijection between solutions $K^-(\lambda)$ and $K^+(\lambda)$ of the left and right reflection equations, as we will show later on.

A way to introduce non-periodic boundary conditions which are compatible with the integrability of the bulk model, was developed in \cite{Sk}. Boundary conditions on the left and right sites of the system will be encoded in the left and right reflection matrices $K^-$ and $K^+$. The compatibility condition between the bulk and the boundary of the system takes the form of the so-called reflection equation \cite{Ch,KuSk}. It is written in the following form for the left reflection matrix acting on the space $\mathbb{C}^2$ at the first site $K^-(\lambda) \in \mathrm{End} (\mathbb{C}^2)$
\begin{equation}
\label{RE}
R_{12}(\lambda - \mu) K^-_1(\lambda) R_{21}(\lambda + \mu) K^-_2(\mu)= 
K^-_2(\mu) R_{12}(\lambda + \mu) K^-_1(\lambda) R_{21}(\lambda - \mu) .
\end{equation}

In complete generality, the compatibility on the right end of the model is encoded in the following dual reflection equation \cite{Sk,FreidelMaillet,KulishSasaki,NADR}
\begin{equation}
\label{rightRE}
A_{12}(\lambda - \mu) K^{+\, t}_1(\lambda) B_{12}(\lambda + \mu) K^{+\, t}_{2}(\mu) = 
K_2^{+\, t}(\mu) C_{12} (\lambda + \mu) K_1^{+\, t}(\lambda) D_{12}(\lambda - \mu) .
\end{equation}
where the matrices $A,B,C,D$ are obtained from the $R$-matrix of the reflection equation \eqref{RE} in the following way
\begin{align}
A_{12}(\lambda)&=\left(R_{12}(\lambda)^{t_{12}}\right)^{-1}=D_{21}(\lambda) , \\
B_{12}(\lambda)&= \left( \left(R_{21}^{t_1}(\lambda)\right)^{-1}\right)^{t_2}=C_{21}(\lambda) ,
\end{align}
explicitly
\begin{equation}
A(\lambda)=\frac{1}{\lambda^2-\eta^2}\left(
       \begin{array}{cccc}
         \lambda-\eta & 0& 0& 0\\
         \lambda\theta & \lambda& -\eta& 0\\
         -\lambda\theta & -\eta& \lambda& 0\\
        \lambda\theta^2 & \lambda\theta& -\lambda\theta& \lambda -\eta\\
       \end{array}
     \right),
\end{equation}
\begin{equation}
B(\lambda)=\frac{1}{\lambda(\lambda+2\eta)}\left(
       \begin{array}{cccc}
         \lambda+\eta & 0& 0& 0\\
         -\lambda\theta & \lambda+2\eta& -\eta& 0\\
         \lambda\theta & -\eta& \lambda+2\eta& 0\\
         -(3\lambda+2\eta)\theta^2 & -\lambda\theta& \lambda\theta & \lambda+\eta\\
       \end{array}
     \right).
\end{equation}

However, due to the property \eqref{weakcross} the dual reflection equation \eqref{rightRE} can be written in the equivalent form
\begin{align}
\label{dRE}
&R_{12}( -\lambda + \mu )K_1^{+}(\lambda) M_2 R_{21}(-\lambda - \mu - 2\eta) M_2^{-1} K_2^{+}(\mu)= \notag\\
&K_2^{+}(\mu) M_1 R_{12}(-\lambda -\mu-2\eta) M_1^{-1} K_1^{+}(\lambda) R_{21}(-\lambda + \mu) .
\end{align}

One can then verify that the mapping
\begin{equation}
\label{bijection}
K^+(\lambda)= K^{-}(- \lambda -\eta) \ M 
\end{equation}
is a bijection between solutions of the reflection equation and the dual reflection equation. After substitution of \eqref{bijection} into the dual reflection equation \eqref{dRE} and using the symmetry property \eqref{Mcom} one gets the reflection equation \eqref{RE} with shifted arguments.

Classification of the solutions is done according to a straightforward but somewhat tedious approach.

First let us note that, obviously, if $K(\lambda)$ is a solution of this equation then so is $f(\lambda)K(\lambda)$. We use this freedom to fix $k_{11}(\lambda)=1$ (and it can be quickly checked that the assumption $k_{11}(\lambda)=0$ leads to a rank 1 noninvertible solution).
We seek the general solution in the form
\begin{equation}
K^-(\lambda)=\left(
       \begin{array}{cc}
         1 & k_{12}(\lambda)\\
         k_{21}(\lambda) & k_{22}(\lambda)\\
       \end{array}
     \right) .
\end{equation}

Replacing this $K$-matrix into \eqref{RE} and writing out the equations for the entries one notices that by adding entries $21$ and $31$ of the resulting $4\times 4$ matrix equality one gets:
\begin{equation}
\label{eq1}
k_{21}(\lambda)\left( k_{22}(\mu)-1\right)=k_{21}(\mu)\left( k_{22}(\lambda)-1\right) .
\end{equation} 
This is a functional equation of the form
\begin{equation}
\label{feq}
f(\lambda) g(\mu) = f(\mu) g(\lambda).
\end{equation}
We recall its general solution. The equation is obviously satisfied if either one of the functions is identically zero. If one of them is not identically zero, they are proportional to each other.

According to this, the solution of \eqref{eq1} is split in two cases:
\begin{enumerate}
\item $k_{22}(\lambda)=\phi k_{21}(\lambda)+1, \quad\phi \in \mathbb{C};$
\item $k_{21}(\lambda)=0.$
\end{enumerate}
We start with \textbf{case $1$} and replace $k_{22}(\lambda)=\phi k_{21}(\lambda)+1$, then the entry $21$ of \eqref{RE} yields
\begin{align}
\label{RE21}
&k_{21}(\lambda)\left(2 \eta \mu +\eta \phi \mu \, k_{21}(\mu)+2 \theta \mu ^2 \, k_{21}(\mu)\right) = \notag \\
&k_{21}(\mu)\left(2 \eta \lambda +\eta \phi \lambda \, k_{21}(\lambda)+2 \theta \lambda ^2 \, k_{21}(\lambda)\right),
\end{align}
which is an algebraic equation for $k_{21}(\lambda)$
\begin{equation}
\label{k21}
k_{21}(\lambda)=\frac{2\eta \lambda}{\xi -\eta \phi \lambda- 2\theta \lambda^2} \ ,
\end{equation}
where $\xi \in \mathbb{C}$ is an arbitrary constant.

So we have now the expression of two elements
\begin{equation}
\label{k22}
k_{22}(\lambda)=1+\frac{2\phi \eta \lambda}{\xi -\eta \phi \lambda- 2\theta \lambda^2} \ .
\end{equation}
Replacing all this in \eqref{RE} we get an equation for $k_{12}(\lambda)$
\begin{equation}
\label{RE12}
\lambda \ k_{12}(\mu)\left( \xi -(\eta \phi +2 \theta \mu) \mu \right) =
\mu \ k_{12}(\lambda) \left( \xi -(\eta \phi +2 \theta \lambda) \lambda \right) ,
\end{equation}
which has the solution
\begin{equation}
\label{k12}
k_{12}(\lambda)=\frac{\psi \lambda}{\xi-\eta \phi \lambda -2\theta \lambda ^2} \ ,
\end{equation}
with arbitrary constant $\psi \in \mathbb{C}$.

Now we turn to \textbf{case $2$} where $k_{21}(\lambda)=0$.
Plugging this assumption into \eqref{RE} leads to
\begin{equation}
\label{RE0}
\lambda \left( 1 + k_{22}(\lambda) \right)\left(k_{22}(\mu) - 1\right)=
\mu \left(1+k_{22}(\mu)\right)\left(k_{22}(\lambda) - 1\right) .
\end{equation}
Here we can assume without loss of generality that $k_{22}\neq 1$.
Then the general solution depends on an arbitrary parameter $\xi$
\begin{equation}
\label{k220}
k_{22}(\lambda)=\frac{\xi +\lambda}{\xi - \lambda} \ .
\end{equation}
Replacing this in the reflection equation leads to
\begin{equation}
\label{RE120}
\lambda k_{12}(\mu)(\mu - \xi ) - \mu k_{12}(\lambda)(\lambda -\xi)=0,
\end{equation}
which has the solution depending on an arbitrary constant $\psi$
\begin{equation}
\label{k120}
k_{12}(\lambda)=\frac{\psi \lambda}{\xi -\lambda} \ .
\end{equation}

Thus, we can identify two families of reflection matrices compatible with a jordanian $R$-matrix in the bulk.
The first family depends on three independent parameters
\begin{equation}
\label{Kfrst}
K^-(\lambda,\psi,\phi,\xi)=\left(\begin{array}{cc} \xi-\phi \eta \, \lambda -2 \theta \, \lambda^2& \psi  \lambda \\
2\eta \lambda & \xi+\phi \eta \, \lambda -2\theta \, \lambda ^2 \end{array} \right) .
\end{equation}
The second family depends only on two independent parameters
\begin{equation}
\label{Ksecond}
K^-(\lambda ,\psi,\xi)= \left(\begin{array}{cc} \xi - \lambda & \psi \lambda\\ 0& \xi + \lambda \end{array} \right) .
\end{equation}

This raw form of the solutions can be transformed after rescaling and redefinition of the parameters into one single family and a more familiar form which reminds one of the general $XXX$ solution
\begin{align}
\label{K-Zol}
K^-(\lambda, \, \xi_-,\phi_-,\psi_-)&= K^-_{XXX}(\lambda, \, \xi_-,\phi_-,\psi_-)  - \phi_- \theta\, \lambda^2 \mathbbm{1}      \notag \\
&= \left( \begin{array}{cc} \xi_- - \lambda- \phi_- \theta \, \lambda ^2& \psi_- \lambda \\
     \eta \phi_- \lambda & \xi_- + \lambda - \phi_- \theta \, \lambda^2 \end{array} \right) .
\end{align}
As it was mentioned earlier, due to relation \eqref{weakcross} the general solution of the dual reflection equation is given by the bijection \eqref{bijection}
\begin{equation}
\label{K+Zol}
K^+(\lambda, \, \xi_+,\phi_+,\psi_+)= K^{-}(-\lambda -\eta , \, \xi_+,\phi_+,\psi_+) \ M .
\end{equation}

Many relations of the XXX spin chain can be obtained from the XXZ-model by simple scaling, i.e. degeneration of the trigonometric functions to the rational ones. It is know that the jordanian deformation of the XXX-chain can also be obtained by a scaling limit with an additional (singular) similarity transformation of the XXZ-model \cite{KulSto1}. To this end one starts from the R-matrix related to the quantum algebra $U_q(sl(2))$  
\begin{equation}
\label{BaxR}
\check{R} (u,q) = u \check{R} (q) - \frac{1}{u} \check{R} ^{-1}(q), \qquad \check{R} (q) = \left(\begin{array}{cccc}q & 0 & 0 & 0 \\ 0 & 0 & 1 & 0 \\0 & 1 & \omega (q) & 0 \\0 & 0 & 0 & q\end{array}\right),
\end{equation}
where $\omega (q) = q - q^{-1}$ and we use the multiplicative parameter $u = \exp \lambda$ in the Yang-Baxter equation \eqref{YBE} and in the reflection equation \eqref{RE}. After the transformation 
\begin{equation}
\label{Jtransf-R}
\check{R} (u,q) \to \mathrm{Ad} J (x) \otimes  J (x) \ \check{R} (u,q),
\end{equation}
with a two-by-two matrix
\begin{equation}
\label{mat-J}
J (x) = \left(\begin{array}{cc}1 & x \\0 & 1\end{array}\right),
\end{equation}
one considers the scaling limit 
\begin{equation}
\label{sclimit}
u = \exp(\epsilon \lambda), \quad q = \exp(\epsilon \eta), \quad x= \frac{\theta}{\eta \epsilon}, \quad \epsilon \to 0.
\end{equation}
The limit of the transformed R-matrix \eqref{Jtransf-R} is proportional to the R-matrix of the twisted Yangian $\mathcal{Y}(sl(2))$ \cite{KhorStolTol}
\begin{equation}
\check{R} (\lambda ,\eta,\theta) = \lambda  \check{R}^{(j)} ( \theta) + \eta \mathbbm{1},
\end{equation}
and hence yielding the deformed XXX-model \cite{KulSto1, He}. 

Being interested in the solution to the reflection equation \eqref{RE} one has to apply the scaling to the K-matrix as well. Although the K-matrix for the XXZ-model is well know \cite{GhosZam,InamiKonno,VeGo2}, the solution corresponding to the R-matrix $\check{R} (u,q)$ (\ref{BaxR}) is different (see e.g. \cite{Doikou,KulMud})
\begin{equation}
\label{DKM-K}
K(u) = \left(\begin{array}{cc} f + u ^2 a & (u^2-u^{-2}) b \\(u^2-u^{-2}) c & f + u ^{-2} a\end{array}\right), 
\end{equation}
with arbitrary parameters $f, a, b, c.$ To get a finite solution after the similarity transformation with the matrix \eqref{mat-J}
\begin{equation}
\label{Jtransf-K}
K(u) \to Ad J(x) \, K(u),
\end{equation}
 one has to scale the parameters in the following way 
\begin{equation}
\label{sclpar}
f = -a + \epsilon \zeta, \quad c = \eta \epsilon c_0, \quad b = \frac{\theta}{\eta \epsilon} (a + \theta c_0) + b_0.
\end{equation}
The limiting solution is then
\begin{equation}
\label{K-Kul}
K(\lambda) = \left(\begin{array}{cc}\zeta + 2\lambda (a +2\theta c_0) & 4\lambda b_0 \\0 & \zeta -2 \lambda (a + 2 \theta c_0)\end{array}\right).
\end{equation}
Comparing it with the solution found above \eqref{Kfrst} one can see that still there are three terms missing. However the scaling approach permits to conclude that the spectra of the deformed model with (restricted) K-matrices (\ref{K-Kul}) coincides with the spectra of the corresponding non-deformed XXX-model.

In order to obtain the complete K-matrix (\ref{K-Zol}) one has to use the following
\begin{equation}
\label{sclpar2}
f = -a + \epsilon \zeta, \quad a = a_0 - \frac{2\theta c}{\eta \epsilon} , \quad b = b_0
+ \frac{\theta}{\eta \epsilon} (a_0 - \frac{\theta}{\eta \epsilon} c) ,
\end{equation}
and also to consider the first three terms in the expansion of $u = \exp(\epsilon \lambda)$. Then obtain 
\begin{equation}
\label{K-compl}
K(\lambda) = \left(\begin{array}{cc}\zeta + 2 a_0 \lambda - \displaystyle\frac{4 \theta c}{\eta} \lambda ^2 & 4b_0 \lambda \\4c\lambda & \zeta - 2 a_0 \lambda - \displaystyle\frac{4 \theta c}{\eta} \lambda ^2\end{array}\right)
\end{equation}
as the limiting K-matrix. Evidently the two K-matrices (\ref{K-Zol}) and  (\ref{K-compl}) coincide if $\zeta = \xi_{-}, \ 2a_0= -1,  \ 4c = \eta \phi _{-},  \ 4b_0= \psi _{-}$.

%
%%%%%%% Construction of open spin chains %%%%%%%%%
%
%

\section{Construction of open spin chains}

To construct deformed integrable open spin chains we follow the method proposed by Sklyanin \cite{Sk}. Taking two arbitrary solutions $K^-(\lambda)$, $K^{+}(\lambda)$
 of the reflection equations \eqref{RE}, \eqref{dRE} we write the open chain transfer matrix as
\begin{eqnarray}
\label{transfer}
t(\lambda)=tr_0\ K^{+}_0(\lambda) \ T_0(\lambda) \ K^-_0(\lambda) \ \widehat{T}_0(\lambda) \ ,
\end{eqnarray}
where the monodromy matrices are given by
\begin{equation}
\label{monodromy}
T_0(\lambda)=R_{0N}(u)\cdots R_{01}(\lambda) , \qquad \widehat{T}_0(\lambda)=R_{10}(\lambda)\cdots R_{N0}(\lambda)  .
\end{equation}
The index $0$ refers to the auxiliary space $\mathbb{C}^2$, while the indices $j= 1, 2, \dots ,N$ refer to the spin \textonehalf  \ spaces at the sites of the chain.

The transfer matrices at different values of the spectral parameter commute in between
\begin{equation}
\label{comutativity}
\left[ t(\lambda) , t(\mu) \right] = 0
\end{equation}
and the open spin chain Hamiltonian can be derived from $t^{\prime}(0) = \frac{d}{d\lambda}t(\lambda)|_{\lambda=0}$. Normalizing the matrix $K^-(0)= \mathbbm{1}$, we write the derivative of the transfer matrix in the form (all arguments set to $0$):
\begin{equation}
\label{transferder}
t^{\prime} (0) \propto \left(tr_0 {K^+_0}^{\prime}\right) + \left(tr_0 K^+_0\right) {K_1^-}^{\prime}+\frac{2}{\eta} tr_0 \left( K^{+}_0 \check{R}_{N0}'\right)  + \frac{2}{\eta}tr_0 {K^+_0}\sum_{j=1}^{N-1} \check{R}_{j,j+1}' .
\end{equation}
From this expression we extract the following Hamiltonian \cite{Sk}
\begin{equation}
\label{Ham}
H=\sum_{j=1}^{N-1} \check{R}_{j,j+1}^{\prime}+\frac{tr_0 K^{+t}_0 \check{R}_{N0}^{\prime}}{tr_0 K^+_0}+\frac{\eta}{2} {K_1^{-}}^{\prime} .
\end{equation}

After replacing the general boundary matrices in this expression we arrive at the following open chain Hamiltonian
\begin{align}
\label{Htotal}
   H &=\sum_{j=1}^{N-1} \frac{1}{2}\left(\sigma_j^x\sigma_{j+1}^x+\sigma_j^y \sigma_{j+1}^y +\sigma_j^z \sigma_{j+1}^z\right)+\theta \left(\sigma_j^+-\sigma_{j+1}^+\right)+\theta^2 \sigma_j^+\sigma_{j+1}^+ \notag\\
&+\frac{\eta}{2}\left( -\sigma_1^z+\psi_- \sigma_1^++\phi_- \sigma_1^-\right) \notag\\
&+\frac{\eta}{2\xi_+} \left( \left(1-\theta \psi_+\right) \sigma_N^z - \left(\theta (2-\theta\phi_+)+\psi_+\right) \sigma_N^+ - \phi_+ \sigma_N^-\right).
\end{align}

This Hamiltonian provides the general choice of open boundary parameters compatible with the integrability of the $XXX_\theta$ model in the bulk. 

Due to the similarity transformations \eqref{Jtransf-R} and \eqref{Jtransf-K}  of the main objects of the QISM the quantization conditions (the Bethe equations) and the spectrum of the periodic Hamiltonian are not changed. Hence, after the limit to the rational XXX-model the spectrum of the periodic deformed model will not change. However, this is not the case with the corresponding eigenvectors, some of them will be transferred into adjoint vectors.

The obtained reflection matrix \eqref{K-Zol} can be put into diagonal form. Its eigenvalues are
\begin{equation}
\label{Ke1,2}
\epsilon _{1,2} (\lambda) = \xi - \phi _- \theta \lambda ^2 \pm \lambda \sqrt{1+\gamma ^2}, \qquad \gamma ^2= \eta \psi _- \phi_ -
\end{equation}
and the matrix of the corresponding eigenvectors $U$ does not depend on $\lambda$
\begin{align}
   K(\lambda) U &= U \ \mathrm{diag} \left( \epsilon _{1} (\lambda) , \epsilon _{2} (\lambda)\right)  , \\
    U &=  \left(\begin{array}{cc}1 & -1 \\ \displaystyle\frac{(x+1)}{\psi _-} & \displaystyle\frac{(x-1)}{\psi _-} \end{array}\right) ,
\end{align}
with $x= \sqrt{1+\gamma ^2}$. However, the approach used e.g. in \cite{AACDFR} to get the Bethe equations defining the parameters of the spin Hamiltonian eigenvectors is not valid in the case of the deformed model. Even the constant $K ^+$ is not identity but the triangular matrix $M$ \eqref{Mmat}, and the R-matrix is not $SL(2)$ invariant.

The spectrum of the free end deformed model with constant reflection matrices $K^-(\lambda) = \mathbbm{1}$ and $K^+(\lambda) = M$ coincides with the spectrum of the free end XXX-spin chain due to the connection of the corresponding Hamiltonians by the similarity transformation. 

%
%%%%%%%%%%% Conclusion %%%%%%%%%%%
%

\section{Conclusion}

We consider XXX-spin chain with non-periodic boundary conditions deformed by a jordanian twist. Solutions $K ^{\pm}(\lambda)$ to the reflection equation and its dual with jordanian R-matrix, are found and they have explicit dependence on the deformation parameter $\theta$. Solution $K ^+(\lambda)$ to the dual reflection equation is obtained due to the generalised crossing symmetry of the jordanian R-matrix and relation $K ^+(\lambda) = K ^-(-\lambda-\eta)M (\theta)$. Thus the transfer matrix $t(\lambda)$ of the model, the generating function of integrals of motion, is obtained. The Hamiltonian of the open spin chain with the general boundary terms at the first and the last sites of the chain is written explicitly. Using these reflection matrices one can study the boundary algebra and possible symmetries of the model.

%
%
%
%\section{Acknowledgments}
%We acknowledge useful discussions with V. ~Tarasov and appreciate the careful reading of the manuscript by the referee.
This work was supported by RFBR grant 07-02-92166-NZNI\_a,\hfil \break 09-01-00504 and the FCT projects
PTDC/MAT/69635 /2006 and \hfil\break PTDC/MAT/099880/2008.

\end{document}